\documentclass{elsart}

\usepackage{amssymb}
\usepackage{epsf,colordvi,color,amsbsy}
\usepackage[dvips]{graphicx}
\begin{document}

\begin{frontmatter}

\title{On the better constraints from the South Pole Telescope Sunyaev-Zel'dovich galaxy clusters survey: a FoM evaluation for the equation of state of Dark Energy}

\author[]{Stefano Sello \corauthref{}}

\corauth[]{stefano.sello@enel.it}

\address{Mathematical and Physical Models, Enel Research, Pisa - Italy}

\begin{abstract}
In a recent article by Benson et al., 2011, the authors show the latest measurements from the South Pole Telescope (SPT) Sunyaev Zel'dovich (SZ) cluster survey to better constrain
some cosmological parameters. In particular, the authors found that adding the SPT cluster data significantly improves the constraints on equation of state of dark energy, $w$, beyond
those found when using measurements of the CMB, supernovae, BAO and the Hubble constant.
The main aim of the present research note is to give a further quantitative estimation of the above better constraints, through the computation of the Figure of Merit (FoM) 
applied to $\Omega_m$ and $w$ plots for the $68\%$ and $95\%$ confidence regions. This allows a better evaluation and a better comparison of the continuous improvements on the cosmological constraints, obtained using new different
cosmological probes and different surveys. 
\end{abstract}
\end{frontmatter}

\section{Introduction}
We recall here some brief considerations of the latest SPT survey, mainly devoted to better constrain some cosmological parameters and, in particular, the equation of state of dark energy, as reported in the reference article by Benson et al., 2011.
Large scale structures in the Universe, like clusters of galaxies, give important information on different cosmological parameters, in particular, the matter density, the amplitude of the matter power spectrum, and the dark energy equation of state, mainly through its effect on the growth of structures.
The cluster abundance measurements are an important tool for testing the standard dark energy model, because they are affected by dark energy properties in a
different way than distance-redshift based tests, such as from type Ia supernovae and baryon acoustic oscillations (Benson et al., 2011).
Recently there has been significant theoretical and experimental progress in the use of galaxies clusters as efficient cosmological probes.
In particular, measurements of the cluster abundance using optical, X-ray, and SZ selection methods have been used to obtain stringent constraints on cosmology and dark energy parameters
Currently, the most precise dark energy constraints from galaxies clusters are obtained from X-ray selected samples which use the X-ray emission from the hot intra-cluster gas as an
index of the total mass in the cluster. We recall here that hot intra-cluster gas also causes a spectral distortion 
in the cosmic microwave background (CMB), a phenomenon known as the Sunyaev-Zel'dovich (SZ) effect (Sunyaev and Zel'dovich, 1972).
The surface brightness of the SZ effect is redshift-independent and largest at mm-wavelengths. 
Thus, a mm-wavelength SZ survey with high angular resolution is expected to provide clean, mass-limited catalogs
out to high redshift, probing the regime where the cluster abundance is most sensitive to dark energy's effect on the growth rate of structure (Carlstrom et al. 2002).
Recently, the first SZ cluster catalogs from three surveys have been released: the South Pole Telescope (SPT, Staniszewski et al. 2009; Vanderlinde et al. 2010;
Williamson et al. 2011), the Atacama Cosmology Telescope (ACT, Marriage et al. 2011, Sehgal et al., 2011), and the Planck satellite (Planck Collaboration et al. 2011).
We recall that the South Pole Telescope is a 10-meter diameter, mm-wavelength telescope, designed to conduct a large-area survey with low
noise and $\sim 1$ arcminute angular resolution and it is sensitive in three bands, at 95, 150, and 220 GHz. The primary goal of the SPT survey is to search
for clusters of galaxies via the SZ effect in a 2500 $deg^2$ survey that was completed in November 2011.
In the article by Benson, et al., 2011, the authors use the SPT-SZ survey with the cluster sample from Vanderlinde et al. (2010), by
incorporating an externally calibrated X-ray observable mass relation and X-ray measurements of the sample in order to present improved cosmological constraints.
More precisely, the authors used a sub-sample of a SZ-selected catalog from the SPT that was described by Vanderlinde et al. 2010
and consisted of 18 clusters at $z > 0.3$ for the cosmological results (for more details see: Benson et al., 2011).

\section{Cosmological constraints}

In addition to the SPT cluster data set,  Benson et al., 2011 incorporate several external cosmological data sets, including measurements of the CMB power spectrum (CMB),
the Hubble constant ($H_0$), Baryon Acoustic Oscillations (BAO), type Ia supernova (SNe Ia), and big bang nucleosynthesis (BBN).
In particular, the authors use measurements of the CMB power spectrum from the seven-year WMAP data release (WMAP7, Larson et al. 2011) and 790 $deg^2$ of sky observed with the SPT (Keisler et al. 2011).
Low-redshift measurements of $H_0$ are from the Hubble Space Telescope (Riess et al. 2011), which are included as a Gaussian prior of $H_0 = 73.8 \pm 2.4$  
$km$ $s^{-1} Mpc^{-1}$. Measurements of the BAO features used SDSS and 2dFGRS data (Percival et al. 2010). The authors included also measurements of the luminosity distances of Type Ia
supernovae (SNe Ia) from the Union2 compilation of 557 SNe (Amanullah et al. 2010), and included their treatment of systematic errors. Finally, the authors considered a BBN
prior from measurements of the abundances of light elements, He and D (Kirkman et al. 2003), which are included as a Gaussian prior of $\Omega_b h^2 = 0.022 \pm 0.002$.
In the paper of Benson et al., 2011, the authors both consider the SPT data constraints for a spatially flat $\Lambda CDM$ cosmological model, and different
extensions, including the dark energy equation of state, $w$, as free parameter, i.e. a wCDM cosmology, a model in which the equation of state of dark energy is
a constant $w$. In this work we are mainly interested to further support and to quantify the improvement in wCDM cosmological constraints
when adding the SPT data to the CMB, BAO, and SNe data sets. In Figure 1, we report a single plot extracted from Figure 4 in Benson et al., 2011, where the authors show
the constraints of the combined CMB+BAO+SNe data set, before and after including the SPT data. "The SPT clusters data most significantly improve the constraints
on $\sigma_8$ and $w$. The combined constraints are $w = -0.973 \pm 0.063$ and $\sigma_8 = 0.793 \pm 0.028$, a factor of $1.25$ and $1.4$ improvement, respectively, over the constraints without clusters. The combined data set also constrains $\Omega_m = 0.273 \pm 0.015$
and $h = 0.697 \pm 0.018$. These constraints are consistent with previous cluster-based results (Vikhlinin et al.
2009; Mantz et al. 2010; Rozo et al. 2010), which used X-ray and optically selected samples of typically lower
redshift clusters." (Benson et al., 2011).
Here we are particulary interested to evaluate and to further quantify these results, using the combined sets with SPT cluster data (SPTCL), in the plane
$(\Omega_m,w)$, relating the mass fraction to the equation of state of dark energy. In order to give a quantitative estimation of the
constraints improvements which can be useful for further comparison using different data sets and different surveys, we computed the Figure
of Merit (FoM) of regions plotted in Figure 1.  
 
\begin{figure}
\resizebox{\hsize}{!}{\includegraphics{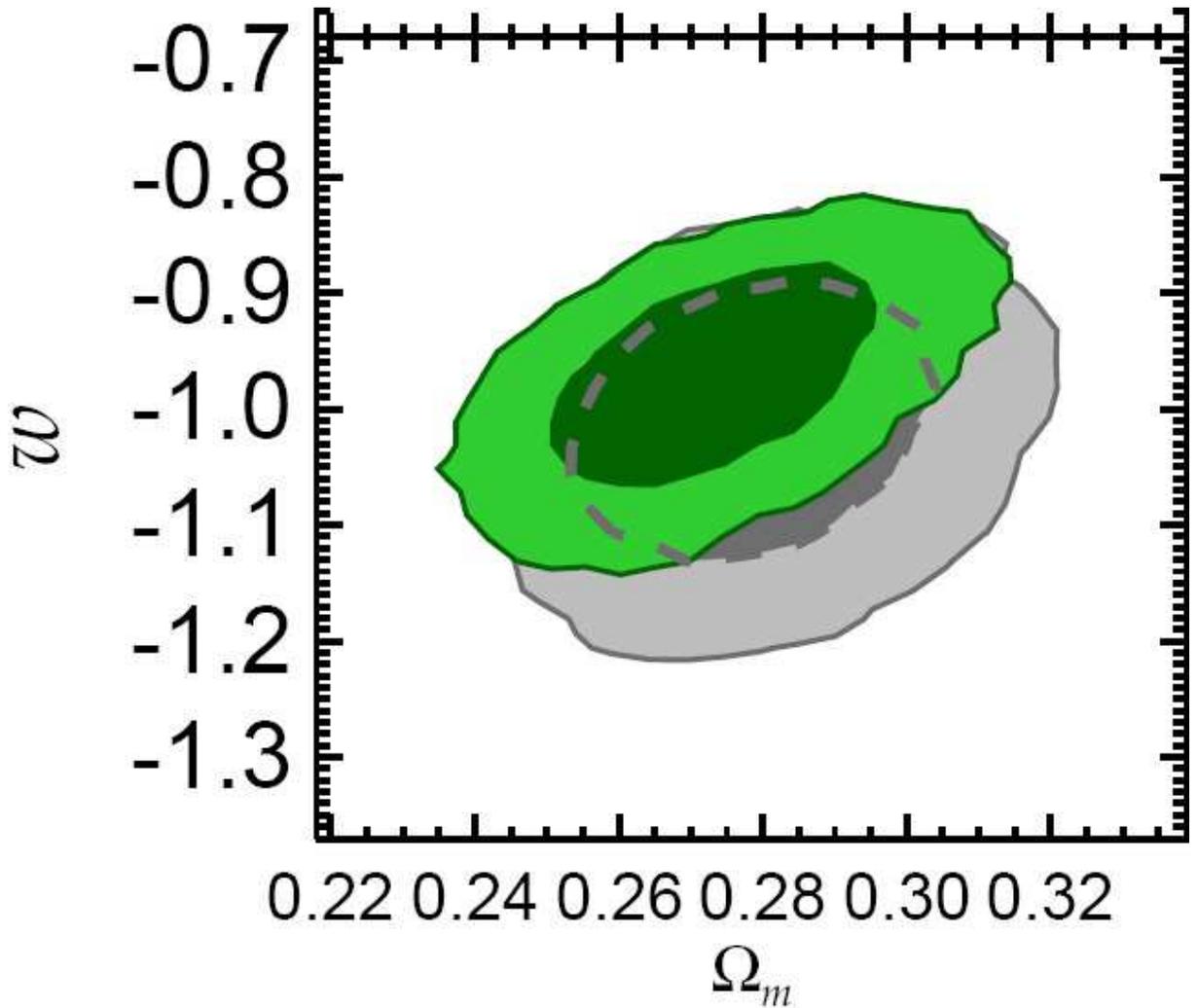}}
 \caption{The constraints on $\Omega_m$, and $w$ assuming a wCDM cosmology.  The contour lines include the two-dimensional marginalized constraints showing the $68\%$ and $95\%$
 confidence regions. We show the constraints for the CMB+BAO+SNe (gray, dashed), and CMB+BAO+SNe+SPTCL (green, solid) data sets. The
SPT clusters data improves the constraints on $w$, by a factor of $1.25$. (From Figure 4 in Benson et al., 2011).}
 \label{fig1}
\end{figure}

\section{FoM evaluation}

In order to make useful comparisons of different dark energy experiments, it is important to choose the appropriate figure of merit (FoM) for dark energy constraints.
The Dark Energy Task Force (DETF) (Albrecht et al., 2006), defined a figure of merit (FoM) that is the inverse of the area enclosed
by the $95\%$ confidence level error ellipse in the selected cosmological parameters plane. We recall that DETF recommended that dark
energy program (Stage III, near-term, medium-cost projects) should be designed to achieve at least a factor of 3 gain over Stage II (ongoing projects) in the FoM, and
that a dark energy program in Stage IV (long-term, high-cost projects, like JDEM, LST, SKA) should be designed to achieve at least a factor of 10 gain over Stage II in the
FoM.
Wang, 2008 proposed a relative generalized FoM given by:

\begin{equation}
    FoM = {1 \over {\sqrt{det Cov(f_1, f_2, ...,f_N)}}} 
\end{equation}
 
where $\{f_i\}$ are the chosen set of cosmological parameters and Cov is the Covariance matrix. For Gaussian distributed errors, FoM is the inverse of the N-dimensional volume enclosed by the $68.3\%$
confidence level contours of all the parameters $\{f_1, f_2, . . . , f_N\}$. This definition is a generalization of the FoM defined by the Dark Energy Task Force (DETF) and has the advantage of being easy to calculate for either real
or simulated data.

We applied the above equation for FoM to plot of Figure 1 to quantify the improvements in the constraints of equation of state of dark energy in the ($\Omega_m$, $w$) plane, for
wCDM model, when we add the SPT clusters data.
In particular, we found that for the $68\%$ confidence regions the FoM value changes from $704.15$ to $1101.6$, considering the constraints without and with clusters, respectively.
This corresponds to an improving factor of $1.56$ or $56.4\%$.
Moreover, for the $95\%$ confidence regions, the FoM value changes from $270.03$ to $387.73$, considering the constraints without and with clusters, respectively.
This corresponds to an improving factor of $1.43$ or $43.6\%$.

As a reference, we note that the discovery of 20 new SNe Ia in the redshift range: $0.623 < z < 1.415$ from the HST Cluster Supernova Survey (studying 25 clusters of galaxies in the redshift range
$0.9 < z <1.5$ with the ACS camera, Suzuki et al., 2011), allowed a new compilation of 580 SNe Ia, called Union2.1, and this improved the constraints on the dark energy density of about $18\%$ for: $1.0 < z < 1.6$.

Another example is given by Wang, 2008b, where there is a plot with the joint confidence contours for $(\Omega_m, \Omega_{\Lambda})$, from an analysis of 307 SNe Ia with and without 69
GRBs (Gamma Ray Bursts), assuming a cosmological constant ($w=-1$). The author shows that the addition of GRB data significantly reduces the uncertainties, and shifts the bestfit parameter
values towards a lower matter density Universe. Here a FoM analyis shows an impressive improving factor of 2.2 (see: Fig.3 in Wang, 2008b).

However, it is important to note that not always adding new samples or new data in a given existing compilation, necessarily improves the constraints for all the cosmological parameters: in fact,
in the two above examples, the inclusion of new supernovae or the addition of GRB data (with or without systematic errors), does not constrains more tightly the equation of state of dark energy,
but here the improving factor remains almost the same (about 1) or even worse, in the sense that we obtain a wider range of possible values for $w$, assuming a wCDM model.
Thus, the better constraints on the equation of state $w$ obtained using the SPTCL data, appear as a valuable result
to improve the knowledge about the nature of dark energy.

The following table summarizes the FoM results.
 
\begin{center}
\begin{tabular}{ | l | l | l | }
\hline
Data set & 68\% & 95\% \\ \hline
CMB+BAO+SNe & 704.15 & 270.03 \\ \hline
CMB+BAO+SNe+SPTCL & 1101.6 & 387.73 \\ \hline
Improving factor & 1.56 & 1.43 \\ \hline
\end{tabular}
\end{center}

The above analysis confirms the significant improvements on constraints for $\Omega_m$ and $w$ parameters gained adding the SPT clusters data to current CMB+BAO+SNe data.
In particular, from Figure 1 we note that the equation of state of dark energy is shifted towards less negative values when we include
the SPT clusters data, currently favoring a cosmological constant model or quintessence models for dark energy. 
It is desirable to perform similar calculations and comparisons when new data will be available from different surveys using different cosmological probes to check the improvements reached in the constraint of cosmological parameters and to verify
the recommendations of the DEFT program.

To complete the FoM analysis, related to equation of state of dark energy, $w$, we also analyzed the gain on constraints for $\sigma_8$ and $w$ parameters (we recall that $\sigma_8$ is the parameter related to the matter fuctuations on 8 Mpc scales at z = 0),
as shown in Benson et al., 2011, using different cosmologies and data sets. In fact, the cluster abundance and the shape of
the cluster mass function depend on $w$ through its effect on the growth of structure, or equivalently the redshift evolution
of $\sigma_8$. Here we assumed, as in Figure 1, the wCDM model and the different data sets considered in Benson et al., 2011.
The following table shows the FoM results for both CMB+BAO+SNe and CMB+BAO+SNe+SPTCL data sets, from Figure 4 in Benson et al., 2011 (see Figure 2).

\begin{figure}
\resizebox{\hsize}{!}{\includegraphics{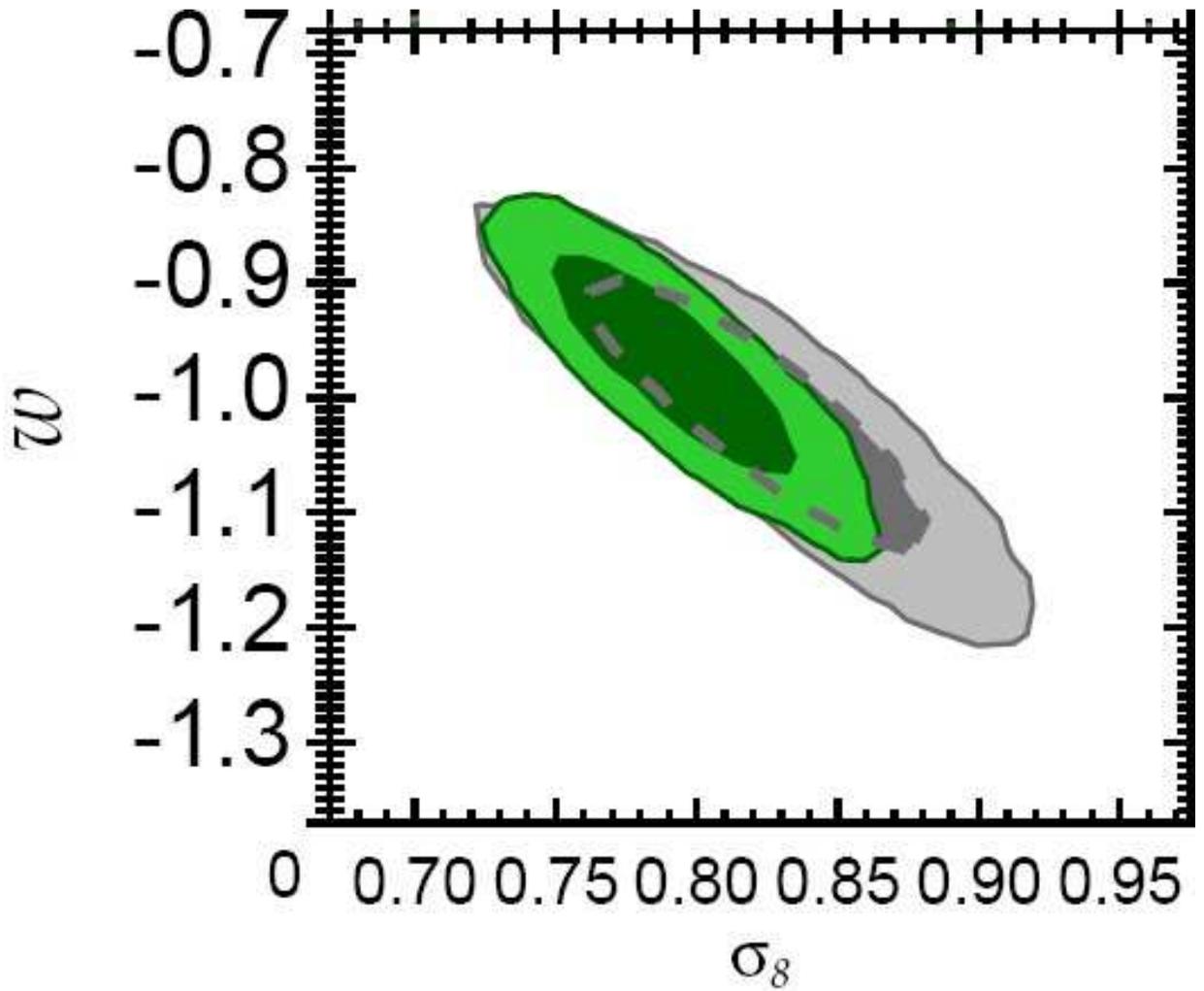}}
 \caption{The constraints on $\sigma_8$, and $w$ assuming a wCDM cosmology.  The contour lines include the two-dimensional marginalized constraints showing the $68\%$ and $95\%$
 confidence regions. We show the constraints for the CMB+BAO+SNe (gray, dashed), and CMB+BAO+SNe+SPTCL (green, solid) data sets. (From Figure 4 in Benson et al., 2011).}
 \label{fig2}
\end{figure}

\begin{center}
\begin{tabular}{ | l | l | l | }
\hline
Data set & 68\% & 95\% \\ \hline
CMB+BAO+SNe & 589.32 & 215.08 \\ \hline
CMB+BAO+SNe+SPTCL & 939.08 & 363.95 \\ \hline
Improving factor & 1.59 & 1.69 \\ \hline
\end{tabular}
\end{center}

Also this analysis shows the significant improvements on constraints for $\sigma_8$ and $w$ parameters gained adding the SPT clusters data to current CMB+BAO+SNe data.

A completely different situation arises from the use of different data sets to constrain the equation of state of dark energy, such as: SPTCL+$H_0$+BBN and CMB. As noted by Benson et al., 2011,
the comparison of the above data sets shows: "The likelihood contours have significant overlap, implying the data are in good agreement. Relative to the CMB, the SPTCL
data tend to disfavor cosmologies with large $\sigma_8$ and more negative $w$." (see Figure 3). However, if we evaluate the overall FoM (i.e. all the possible value-combinations, not only the extreme values)
for both the data sets and for both the confidence regions, we obtain significant lower tight constraints (improving factor $<1$) in the plane: $(\sigma_8, w)$.

\begin{figure}
\resizebox{\hsize}{!}{\includegraphics{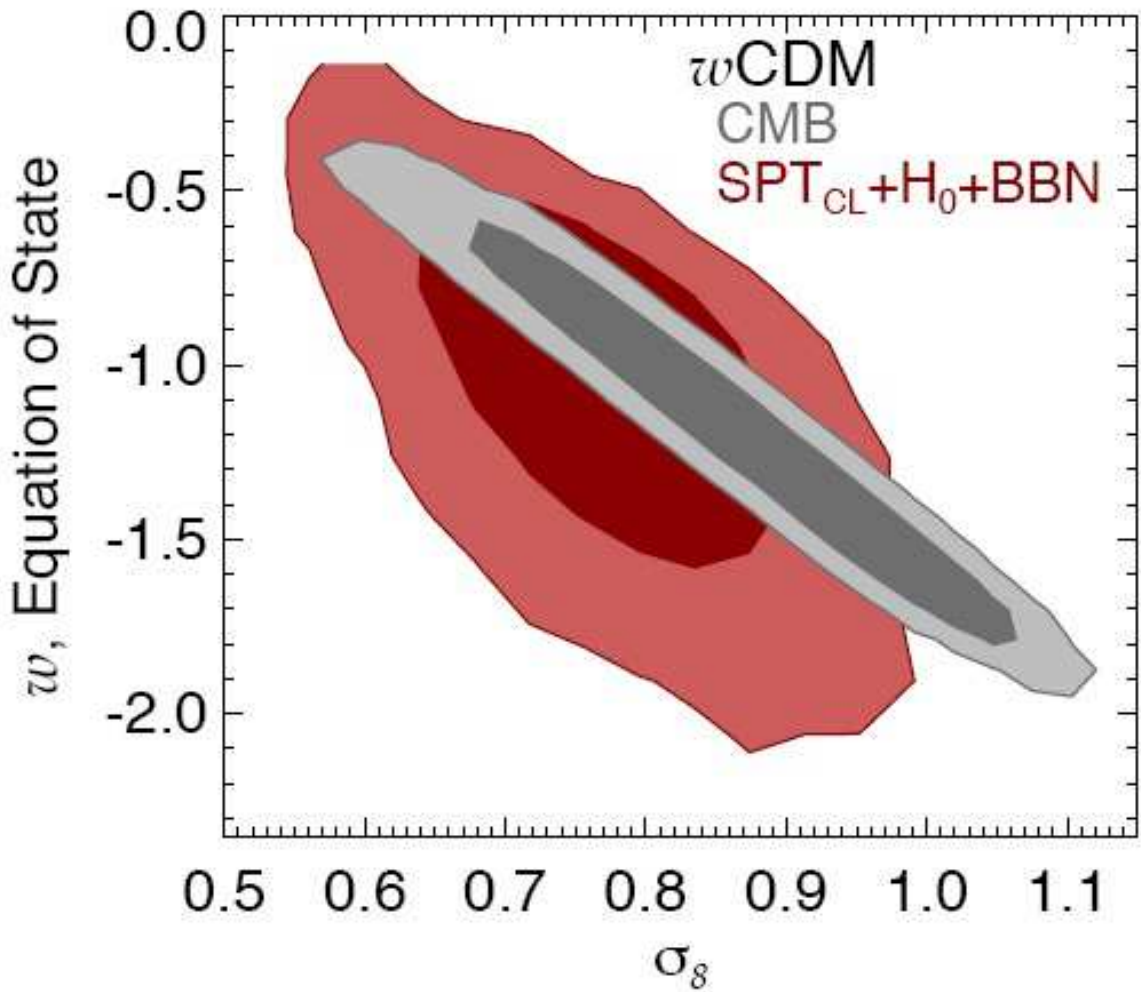}}
 \caption{The constraints on $\sigma_8$, and $w$ assuming a wCDM cosmology.  The contour lines include the two-dimensional marginalized constraints showing the $68\%$ and $95\%$
 confidence regions. We show the constraints for the SPTCL+$H_0$+BBN (red) and CMB (gray) data sets. (From Figure 3 in Benson et al., 2011).}
 \label{fig3}
\end{figure}

The following table shows the related FoM values.

\begin{center}
\begin{tabular}{ | l | l | l | }
\hline
Data set & 68\% & 95\% \\ \hline
CMB & 81.19 & 36.32 \\ \hline
SPTCL+$H_0$+BBN & 37.04 & 12.64 \\ \hline
Improving factor & 0.46 & 0.35 \\ \hline
\end{tabular}
\end{center}

As a comparison, previous constraints in the plane $(\sigma_8, w)$, obtained adding to current CMB data the ACT-SZ sample of 9 optically-confirmed high-mass clusters comprising the high-significance
end of the total cluster sample identified in 455 $deg^2$ of sky surveyed during 2008 at 148 GHz (Sehgal et al., 2011), show an outstanding improving factor of 2.56, at $68\%$ confidence regions, when
compared with CMB WMAP7 data alone and using a fixed fiducial mass scaling relation; whereas we obtain a slight improving factor of 1.16, when allowing the parameters of the mass scaling relation to vary
(for more details, see Figure 6 in Sehgal et al., 2011).

As well noted in Benson et al., 2011, regarding the prospects for further improvement, "...applying this (mass) calibration to the full 2500 $deg^2$
SPTCL+$H_0$+BBN data set, we should constrain $w$ with an accuracy of $\sim 8\%$, or a factor of $\sim 4.5$ tighter than
the current SPTCL+$H_0$+BBN constraints. This improved constraint would be comparable to the current constraints from the CMB+BAO+SNe data, and would
be an independent systematic test of the standard dark energy paradigm by measuring the effect of dark energy on the growth of structure."
From our FoM analysis follows that, in order to reach a comparable constraint level in the overall plane: $(\sigma_8, w)$, the current SPTCL+$H_0$+BBN data set
needs to improve by a factor $>25$, if compared to current CMB+BAO+SNe+SPTCL data (actually a factor $\sim 5.6$ for the allowable range of $w$ only, at $68\%$ confidence region).

\section{Conclusions}

Using the latest measurements from the South Pole Telescope (SPT) Sunyaev Zel'dovich (SZ) cluster survey, Benson et al., 2011, found that adding the SPT cluster data significantly improves the constraints on equation of state of dark energy, $w$, beyond
those found when using measurements of the CMB, supernovae, BAO and the Hubble constant.
In this research note we performed a further quantitative estimation of these better constraints, through the computation of the Figure of Merit (FoM) 
applied to $\Omega_m$ and $w$ plots for the $68\%$ and $95\%$ confidence regions. Our analysis confirms the significant improvements on constraints for $\Omega_m$, and $w$ parameters
gained adding the SPT clusters data to current CMB+BAO+SNe data. The improving factors are $1.56$ and $1.43$ for confidence regions at $68\%$ and $95\%$, respectively.
Similar conclusions are obtained considering the constraints on the plane: $(\sigma_8, w)$.
This allows us to better evaluate and to better compare the current and future improvements obtained on the cosmological constraints,
using both different cosmological probes and different surveys.

\section{References}

Albrecht, A., et al., 2006, Report of the Dark Energy Task Force, astroph/0609591

Amanullah, R., et al. 2010, ApJ, 716, 712

Benson, B.A., et al., 2011, astroph.CO/1112.5435v1

Carlstrom, J. E., Holder, G. P., and Reese, E. D. 2002, ARAA,40, 643

Keisler, R., et al. 2011, ApJ, 743, 28

Kirkman, et al., 2003, ApJS, 149, 1

Komatsu, E., et al. 2009, ApJS, 180, 330;  2011, ApJS, 192, 18

Larson, D., et al. 2011, ApJS, 192, 16

Linder, E. V. and Jenkins, A. 2003, MNRAS, 346, 573

Mantz, A., Allen, S. W., Rapetti, D., and Ebeling, H. 2010, MNRAS, 406, 1759

Marriage, T. A., et al. 2011, ApJ, 731, 100

Percival, W. J., et al. 2010, MNRAS, 401, 2148

Planck Collaboration, et al. 2011, AA, 536, A8

Riess, A. G., et al. 2011, ApJ, 730, 119

Rozo, E., et al. 2010, ApJ, 708, 645

Sehgal et al., 2011, ApJ, 732, 44

Staniszewski, Z., et al. 2009, ApJ, 701, 32

Sunyaev, R. A., and Zel'dovich, Y. B., (1972), Comments on Astrophysics and Space Physics, 4, 173

Suzuki, N. et al., 2011, astroph.CO/1105.3470v1

Vanderlinde, K., et al. 2010, ApJ, 722, 1180

Vikhlinin, A., et al. 2009, ApJ, 692, 1033;  2009, ApJ, 692, 1060

Wang, Y., 2008, Phys. Rev. D, 77, 123525

Wang, Y., 2008b, astroph/0809.0657v3

Williamson, R., et al. 2011, ApJ, 738, 139

\end{document}